\documentclass{emulateapj}

\usepackage[usenames]{color}

\shorttitle{Chandra Observations of 3C 445}
\shortauthors{Perlman et al.}

\begin{document}

\title{Chandra Observations of the Radio Galaxy 3C 445 and the Hotspot  X-ray Emission
Mechanism}

\author{Eric S. Perlman\altaffilmark{1}, Markos Georganopoulos\altaffilmark{2,3},  Emily M. May \altaffilmark{4,5}, Demosthenes Kazanas\altaffilmark{3}}

\altaffiltext{1}{Department of Physics and Space Sciences, Florida Institute 
of Technology, 150 W. University Blvd., Melbourne, FL 32901}

\altaffiltext{2}{Department of Physics, 
University of Maryland, Baltimore County, 1000 Hilltop Circle, Baltimore, MD 
21250}

\altaffiltext{3}{Laboratory for High Energy Astrophysics, NASA Goddard Space 
Flight Center, Codel 661, Greenbelt, MD 20771}

\altaffiltext{4}{Department of Physics and Astronomy, University of Wyoming, WY 82071}

\altaffiltext{5}{Southeastern Association for Research in Astronomy (SARA)
NSF-REU Summer Intern at Florida Institute of Technology}

\begin{abstract}
  
We present new {\it Chandra} observations of the radio galaxy 3C 445, centered
on its southern radio hotspot.  Our observations detect  X-ray emission
displaced upstream and to the west of the radio-optical hotspot. Attempting to
reproduce both the observed spectral energy distribution (SED) and the
displacement, excludes all one zone models. Modeling of the radio-optical
hotspot spectrum suggests that the electron distribution has a low energy cutoff
or break approximately at the proton rest mass energy.  The X-rays could be due
to external Compton scattering of the cosmic microwave background (EC/CMB)
coming from the fast (Lorentz factor $\Gamma\approx 4$) part of a decelerating
flow, but this requires a small angle between the jet velocity and the
observer's line of sight ($\theta\approx 14^{\circ}$). Alternatively, the X-ray
emission can be synchrotron from a separate population of electrons.  This last
interpretation does not  require the X-ray emission to be beamed.

 \end{abstract}


\section{Introduction \label{intro}}


The hotspots of powerful Fanaroff-Rilley type II [FRII;  Fanaroff \& Riley 1974]
radio galaxies and quasars are the locations where the jets of these sources,
after propagating for distances up to $\sim 1$ Mpc, terminate in a collision
with the intergalactic medium. The optical emission observed in several
hotspots  suggests  that at least a fraction of the electrons that goes through
the shock(s) formed at the termination  of the jets undergoes efficient particle
acceleration \citep{heavens87, meisenheimer89, prieto02}.  {\sl Chandra} results
show that while in some sources (e.g. Cygnus A; Wilson et al. 2000) the hotspot
X-ray emission is consistent with synchrotron-self Compton radiation from
relativistic electrons  in energy equipartition with the magnetic field (SSCE),
in other sources (e.g. the hotspot on the jet side of Pictor A; Wilson et al.
2001,  Hardcastle \& Croston 2005, Migliori et al. 2007, 
Tingay et al. 2008) the X-ray emission 
is at a much higher level  (by up to a factor of $ \sim
1000$). The nature of this anomalously bright (significantly brighter than SSCE) X-ray hotspot emission  remains a matter of active discussion 
in the literature, being connected to the issue of particle acceleration
efficiency  and jet power [for two recent  reviews see Harris \& Krawczynski
(2006) and Worall (2009)] and extending to a similar and possibly related issue
for the knots of powerful jets (e.g. Kataoka \& Stawarz 2005).

Two different considerations appear to be relevant to the collective properties
of hotspot X-ray emission, relativistic beaming and hotspot  luminosity;
however, it is not as yet clear how exactly  these may be combined to reproduce
the  observed phenomenology. Based on early {\sl Chandra} results, suggesting
that in most cases the anomalously bright X-ray hotspots were seen at the
approaching  jet of  sources with jets forming relatively small angles to the
observer's line of sight, Georganopoulos \& Kazanas (2003) proposed that  the
X-ray emission is  beamed, and that the plasma in the hotspot is relativistic
and decelerating from a bulk Lorentz factor $\Gamma\sim 2-3$ to velocities that
match the subrelativistic advance speed of lobes  ($u/c\approx 0.1$; Arshakian
\& Longair 2001). In this scenario the X-ray emission  is mostly due to upstream
Compton (UC) scattering of electrons in the upstream fast ($\Gamma\sim 2-3$)
part of the flow,  inverse-Compton scattering the  synchrotron photons produced
downstream. The relevance of beaming is also supported by a study of  quasar
hotspots by Tavecchio et al. (2005) that concluded that the anomalously bright
X-ray emission is indeed found mostly on the hotspot corresponding to the
approaching jet side, although these authors favor the cosmic microwave
background as a source of seed photons. On the other hand, Hardcastle et al.
(2004) find only weak evidence that the anomalously X-ray bright hotspots are
more frequently found at the termination of the approaching jet. 

The second consideration, namely the hotspot luminosity, was introduced by
\cite{brunetti03}, who found that the synchrotron emission seen at radio 
energies extends to the optical for lower power sources like 3C 445
\citep{prieto02}  but cuts off before the optical regime for powerful sources
like Cygnus A \citep{wilson00}.  They explained this as a consequence of 
radiative losses increasing with hotspot luminosity. The luminosity
range over which this decrease of the synchrotron peak frequency with increasing
luminosity is observed, was extended to powerful hotspots by \cite{cheung05}.
Based on an extensive sample of hotspot multiwavelength data, 
\cite{hardcastle04} argued for the relevance of the hotspot luminosity to the
X-ray emission, by showing that  hotspots with X-ray luminosity much higher than
that predicted by SSCE 
were usually of  low luminosity, in contrast to more powerful
hotspots.  
However, instead of UC, \cite{hardcastle04} favored synchrotron emission from 
an altogether separate
electron population as the source of the anomalously bright X-ray emission.
Other workers have also come to the conclusion that synchrotron radiation 
from a second electron population is the
most likely emission mechanism for this component (e.g., in the case of Pic A,
see Fan et al. 2008, Tingay et al. 2008), partly as a result of observing 
compact radio hotspots with the VLBA. 

Synchrotron radiation, but  from a single electron population, is indeed the
X-ray emission mechanism for the jets of low power FR I radio galaxies, as seen
from their single component radio-optical-X-ray spectra (e.g. Perlman \& Wilson
2005 for the jet of M 87). This turns out to be the case also for some of the
weakest hostspots of FR II radio galaxies: the northern hotspot of 3C 390.3
\citep{hardcastle07}, the northern hotspots of 3C 33 \citep{kraft07}, the
eastern hotspots of 3C 403 \citep{kraft07} and both hotspots of $0836+299$
\citep{tavecchio05}. It comes as no surprise that in these sources  their X-ray 
emission is much higher than the anticipated SSC flux, since the X-ray emission
is indeed the continuation of the radio-optical synchrotron spectrum to higher
energies. We therefore do not consider sources that exhibit a single spectral
component from radio to X-ray emission to be part of the family  of hotspots 
exhibiting anomalously high X-ray emission.

An additional handle in understanding the X-ray emission process of the
anomalously X-ray bright hotspots (those for which the X-rays (i) cannot be a
continuation of the synchrotron spectrum and (ii) are significantly brighter
than predicted by SSCE) can come from spatial displacements between the 
emission at different frequencies. In a handful of these hotspots, 
displacements between the radio and the X-ray hotspot emission have
been observed, with the X-rays being upstream of the radio: 3C 351
\citep{hardcastle02}, 4C 74.26 \citep{erlund07}, 3C 390.3 and  3C 227
\citep{hardcastle07}, 3C 321 \citep{evans08}, 3C 353 \citep{kataoka08}.   
{\sl So far no
displacements have been observed in the hotspots of sources for which their 
X-ray emission is consistent with SSCE.} 
An additional important characteristic is that when optical-IR emission is
detected from these hotspots (as in 3C227, 3C 390.3, 3C 351) it coincides with
or is shifted somewhat upstream of the radio, to a location downstream
of the X-rays. Beaming in the hotspots that exhibit displacements can be a
relevant but not a dominant influence, because, while in three of them the
hotspots are in the approaching jet that points toward us (4C  74.26, 3C 227,
3C 351),   in two others (3C 321 and 3C 353), the jets are believed to be
close to the plane of the sky, while in superluminal 3C 390.3 the  hotspot
exhibiting the X-ray radio displacement is  at the termination of the counter
jet. Similarly, a low hotspot power does not seem to be strictly required,
since the hotspot of 3C 351 is rather powerful and still exhibits the
displacements mentioned above. 
  
In this paper we present {\sl Chandra} X-ray observations and discuss the X-ray
emission mechanism of the hotspots of 3C 445, a broad line FR II radio galaxy at
redshift $z=0.0562$  \citep{eracleous94}, which for the standard cosmology
($H_0=71 $ km s$^{-1}$ Mpc$^{-1}$, $\Omega_\Lambda=0.73$ and $\Omega_M=0.27$)
corresponds to a luminosity distance of $247.7 $ Mpc. This is a promising source
for constraining the hotspot X-ray emission mechanism with high resolution 
{\sl Chandra} observations.  Both its southern and northern  hotspots have been
detected in the radio and in the near-IR/optical, with the southern hotspot
being $\sim 3$ times brighter than the northern and having multiple sites of
synchrotron optical emission, a manifestation of ongoing particle acceleration
\citep{prieto02,mack09}.  The hotspot - counter hotspot 
luminosity difference can be intrinsic, or due to mild beaming with the southern
hotspot being on the approaching jet side. 

In the first case, adopting the suggestion of \cite{hardcastle04}, we expect
X-ray emission brighter than SSCE from both hotspots, because this is a source
with a modest extended power at  $178$ MHz, $P_{178}=3.0 \times 10^{25}$ W
Hz$^{-1}$ sr$^{-1}$  \citep{hardcastle98}, as well as a low $5$ GHz  luminosity from
both hotspots  ($L_{5 GHz}=3.4 \times 10^{22}$ W Hz$^{-1}$ sr$^{-1}$ for the
northern hotspot and $L_{5 GHz}=7.9 \times 10^{22}$ W Hz$^{-1}$ sr$^{-1}$ for the
southern hotspot \citep{mack09}). In this scenario, the X-ray emission from the weaker
northern hotspot is expected to be higher than predicted by 
the X-ray to radio ratio  
from the brighter southern hotspot.  This model makes no prediction
regarding offsets between emission in different bands.

In the second case, we expect that due to beaming, the presumed 
approaching-jet side,
southern hotspot will have more pronounced X-ray emission.  In the context of
a relativistic decelerating hot spot flow  \citep{georganopoulos03,
georganopoulos04},  we also expect to observe  offsets, with the X-rays peaking
upstream of the radio. If, in addition, distributed particle acceleration takes
place (as suggested by Prieto et al. 2002), the optical and radio emission will
peak at the same location (see Figure 3 of Georganopoulos \& Kazanas 2004).


3C 445 has been the subject of several X-ray observations.  {\it GINGA} 
observations were reported by Pounds (1990), while {\it ASCA} observations 
were reported by Sambruna et al. (1998).  Both observations reported an
absorbed Seyfert
nucleus with a hard X-ray continuum.  
{\it XMM-Newton} observations of 3C 445 were previously published by
\cite{sambruna07} and  \cite{grandi07},  but in those data a partial window
configuration was chosen for the MOS and pn that caused both the northern and
southern hotspots to be at locations where the CCDs were not read out.  
This work thus represents the first
study of extended X-ray emission from this object, and we report the discovery
of X-ray emission from  the southern hotspot, as well as a meaningful X-ray flux
upper limit from the northern hotspot. In \S 2 we discuss the {\it Chandra} 
observations, as well as the data reduction and analysis procedures.  In \S 3 we
present and contrast the morphology seen in each band for the southern hotspot.
In \S 4 we  discuss the  X-ray emission mechanism of the southern hotspot,
present our conclusions and suggest directions for future work.

\section{Observations and Data Reduction \label{obs}}

3C 445 was observed with the Advanced CCD Imaging Spectrometer (ACIS) on the
Chandra X-ray Observatory on 18 October 2007  for 50ks. The ACIS-S configuration
was used due to its linear alignment which allowed all emission
regions (in particular both hotspots) to be included in the observation.
The northern and southern hot spots are at projected distances
of $\sim$ 300 kpc (more than 5 arcminutes in angular distance) from the core of
the galaxy. However, because of the greater
surface brightness of the southern hotspot (seen in the radio and near-IR) we
decided to center it in the  observations to maximize our sensitivity and
angular resolution in this region.

The {\it Chandra} observations were reduced in the Chandra Interactive
Analysis of Observations (CIAO) software package.  Standard recipes (i.e., the
{\it science threads}) were followed for imaging spectroscopy of extended 
sources as well as data preparation and filtering.  No significant flare events
were seen during the observation, so all data could be used in the analysis. 
We extracted spectra for the southern hotspot and nucleus, and created an
exposure map to allow a search for emission from other  source regions.   All
X-ray spectra were fitted in XSPEC.  This process is discussed in \S 3.  

Deep optical and near-infrared observations of the northern and southern hot
spots were obtained by \cite{prieto02} with the Very Large Telescope (VLT) of
the European Southern Observatory (ESO) and the Infrared Spectrometer and Array
Camera (ISAAC) in the K$_{S}$ (2.2 $\mu$m), H (1.7 $\mu$m), J$_{S}$ (1.2
$\mu$m), and I (0.9 $\mu$m) bands; deeper images in these same bands plus the R
(0.7 $\mu$m) and B (0.45 $\mu$m) bands were obtained by \cite{mack09}.    We
refer the reader to those papers for a  discussion of their data reduction
procedures.  We obtained near-IR and optical fluxes from their  paper, as well
as from \cite{mack09}.   We obtained their {\it VLT} $J_S$-band image  
for use in this paper. 

We extracted radio data for 3C 445 from the NRAO data archives. 3C 445 was
observed with the NRAO {\it Very Large Array} on 09 September  2002, at both 8.5
and 5.0 GHz.  The {\it VLA} was in the B configuration,  yielding a clean beam
(resolution element) of 0.82 $\times$ 0.72 arcseconds in PA 77.44$^\circ$. 
Those data were originally analyzed by \cite{brunetti03}; we reanalyze them in
this paper.  Data reduction was done in {\it AIPS}, using standard procedures.
We also obtained radio fluxes in other bands from \cite {brunetti03}. 

In registering the three datasets to a common frame of 
reference, we assumed the VLA map to be the fiducial, adhering to the
usual IAU standard.  The VLT image
was registered to this frame by comparing to Palomar Sky Survey and USNO-A2.0 
data.  The radio galaxy itself could not be used
for this  procedure, as it was out of the field of view of both the VLT and VLA
images.  It was therefore necessary to use stars in the field
to do the registration. 
We then  checked the alignment of the radio and optical images by
overplotting the two, in the process reproducing the Prieto et al. (2002)
overlay. Following this, the 1$\sigma$ error in the positions from the VLT 
image are $\pm 0.2''$  in RA and Dec (see e.g., Deutsch 1999) relative
to either the radio or X-ray images, while those in the X-ray image are
$\pm 0.4''$ {\footnote{see the {\it Chandra} Science Thread on Astrometry, 
http://cxc.harvard.edu/cal/ASPECT/celmon}}, relative to either the radio 
or optical. 

\section{Results \label{res}}


Both the northern and southern hotspots of 3C 445 are seen in multiple
optical/near-IR bands (Prieto et al. 2002, Mack et al. 2009).  However,  X-ray
emission was seen from the southern hotspot only.  In Table 1, we  list all
fluxes for both hotspots, using all available data.  In Figure 1, we show 
the X-ray, radio and near-IR images of the southern radio hotspot. The 
radio and near-IR images are shown both at the pixel scale of the {\it Chandra}
data (i.e., $0.492''$/pix, middle and bottom left panels) as well as at the 
finer pixel scale of the VLA image (i.e., $0.220''$/pix, middle and bottom
right panels).
Our {\sl Chandra} observations detect about 200 counts from the southern
hotspot.  The X-ray emission extends along a $6''$  region, extending very 
nearly east-west and peaking near the middle or perhaps slightly to the west
of its center point.

\begin{deluxetable}{ccccl}
\tablecaption{Hotspot fluxes}
\tablehead{
\colhead{Feature} & \colhead {Telescope} & \colhead{Flux ($\mu$Jy)} & 
\colhead{Frequency (Hz)} & \colhead{Source}}
\startdata 

3C 445 N  & VLA  &  4000000           &   74$\times 10^8$ & 4\\
3C 445 N  & VLA  &  1400000           &   330$\times 10^8$ & 4\\
3C 445 N  & VLA  &	160000		 &  1.4$\times 10^9$  &  2 \\
3C 445 N  & VLA  &	58000		 &  4.8$\times 10^9$  &  2 \\
3C 445 N  & VLA  &	35000		 &  8.4$\times 10^9$  &  2 \\
3C 445 N  & VLT  &     $7.2\pm 2.2$      &  1.38$\times 10^{14}$ & 2 \\
3C 445 N  & VLT  &     $<2.9$            &  1.81$\times 10^{14}$ & 2  \\
3C 445 N  & VLT  &     $2.4\pm 0.7  $    &  2.47$\times 10^{14}$ & 2 \\
3C 445 N  &  {\it Chandra} & $<2.60\times 10^{-4}$  & 1.21$\times 10^{18}$ & 3 \\
\hline
3C 445 S  & VLA  &  7900000             &   74$\times 10^8$ & 4\\
3C 445 S  & VLA  &  2000000           &   330$\times 10^8$ & 4\\
3C 445 S  & VLA  &	520000		 &  1.4$\times 10^9$  & 1  \\
3C 445 S  &  VLA &	135000		&   4.8$\times 10^9$  & 1  \\
3C 445 S  & VLA  &       81000		&   8.4$\times 10^9$  & 1  \\
3C 445 S  & VLT   &    $16.5 \pm 1.5 $   &   1.38$\times 10^{14}$ & 2  \\
3C 445 S  & VLT   &    $15.2 \pm 3.0 $   &   1.81$\times 10^{14}$ & 2  \\
3C 445 S  & VLT   &    $13.6 \pm 1.4 $   &   2.47$\times 10^{14}$ & 2  \\
3C 445 S  & VLT   &    $5.65 \pm 0.57$   &   3.33$\times 10^{14}$ & 2   \\
3C 445 S  & VLT   &    $4.56 \pm 0.90$   &   4.29$\times 10^{14}$ & 2  \\
3C 445 S  & VLT   &    $2.95 \pm 0.60$   &   6.82$\times 10^{14}$ & 2  \\
3C 445 S  & VLT   &    $0.95 \pm 0.19$  &    8.33$\times 10^{14}$ & 2  \\
3C 445 S  &  {\it Chandra} & $9.38\times 10^{-4} $&     1.21$\times 10^{18}$ & 3  \\
\enddata
\tablerefs {(1) Brunetti et al. (2003); (2) Mack et al. (2009); 
(3) This paper (4) Kassim et al. (2007) and Kassim, private communication}
\end{deluxetable}

There are clear differences between the 
morphologies seen in the near-IR, X-ray and radio (Figure 1).  In contrast to
the X-ray emission, in the near-IR and radio we see 
a concave arc, which appears to contain the X-ray emission in its hollow part.
This can be seen in Figure 1's left-hand panels, which show the data 
at a common pixel scale. There also appear to
be offsets between the position of the X-ray emission component  and
those seen in the near-IR and radio. Looking at higher resolution (middle 
right and bottom right panels), 
the near-IR emission breaks up  into three discrete
regions, which we term NIR 1-3 in east-west order, shown also in 
Mack et al. (2009).   The brightest near-IR emission is located within
region NIR 1.  
The radio emission peak is also near this position, and the
higher-resolution VLA data of Mack et al. (2009) show that the radio emission
shows a broad plateau that includes the entire maximum region of NIR 1.
\begin{figure*}
\epsscale{1.1}
\plotone{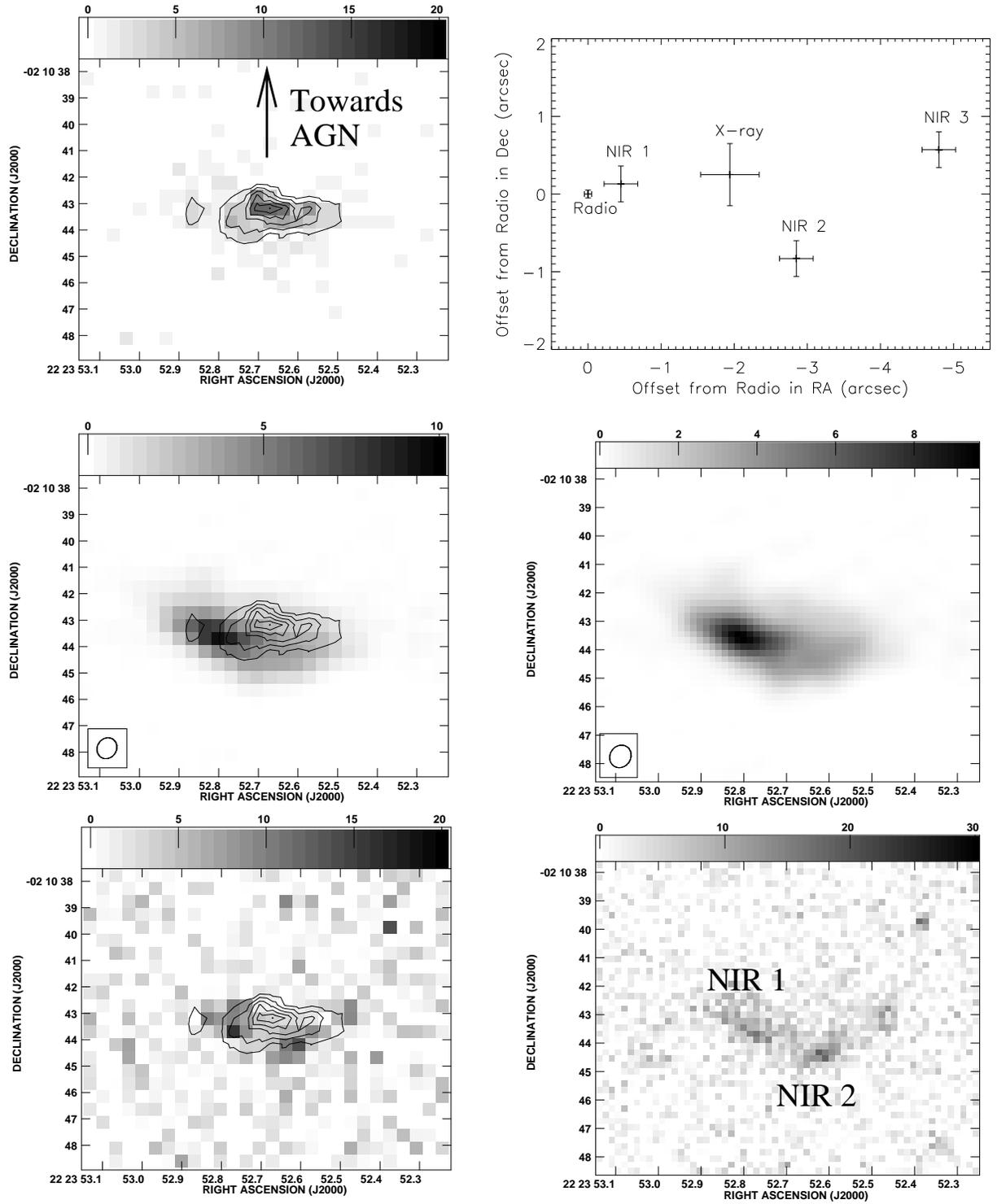}
\caption{The southern hotspot of 3C 445, as seen in the X-rays (top left 
panel), radio (middle panels), and near IR  (bottom panels).     
The images at left were
resampled to 0.492
$''$/pix, while the middle right and bottom right panels are at  
0.220$''$/pix.  
The contours were taken from the {\it Chandra} image,
smoothed with a 1-pixel (FWHM) Gaussian.   Contours are plotted at 2, 4, 6...
counts/pixel.
The top right panel shows the 
centroid of the emission components in each band, plotted with
error bars relative to the radio frame. See \S 3 for discussion.}
\label{fig:montage}
\end{figure*}

We have attempted to quantify the displacements between components in
different bands by 
fitting elliptical Gaussians to each major component within AIPS, using
the task JMFIT.  In order to utilize the full resolution of our data, we
did this measurement at the native resolution in each band, using small 
boxes to zero in on the visible maxima.
These positions are reported in Table 2, and are also
compared in the top right-hand panel of Figure 1. In the case
of the near-IR emission, we report three centroids, one for each of the
three regions seen in that image (Figure 1); these have been labeled 1-3,
in order from east to west.  We report in Table 2 with parentheses 
the internal errors from 
JMFIT (estimated at 0.2 pixels where smaller values were reported).  However,
for cross-comparison between bands, we  must emphasize that the errors are
dominated by the uncertainty in registration between bands, i.e., 
$\sim 0.2-0.4''$, as detailed in \S 2.  Note that these are errors in
cross-comparison -- i.e., they do not constitute errors on each individual
position but rather get added only once to the internal errors reported in 
Table 2.     
As can be seen, the displacement between the X-ray and radio peak is
significant at $>3 \sigma$, while the displacements between the X-ray peak
and those of NIR 1 and NIR 2 are at the 2.4-3 $\sigma$ level.

\begin{deluxetable}{cllc }
\tablecaption{Positions of Emission Regions in Southern Hotspot}
\tablehead{
\colhead{Band} & \colhead {RA (J2000)} & \colhead{Dec} & 
\colhead{Delta(radio)}  }
\startdata 

X-ray & 22 23 52.67~~(0.01) & -02 10 43.29~~(0.05) & (-1.94,+0.25) \\
NIR 1 & 22 23 52.77~~(0.01) & -02 10 43.67~~(0.12) & (-0.45,+0.13)  \\ 
NIR 2 & 22 23 52.61~~(0.01) & -02 10 44.37~~(0.12) &  (-2.85,-0.83) \\
NIR 3 & 22 23 52.48~~(0.01) & -02 10 42.97~~(0.22) & (-4.80,+0.57)  \\
Radio & 22 23 52.80~~(0.01) & -02 10 43.54~~(0.02) & (0,0)  \\

\enddata
\end{deluxetable}

Our {\it Chandra} data include the position of the northern hotspot, which
was detected for the first time in the near-IR by Mack et al. (2009), who
describe
its radio and optical morphology. We do not detect it in the {\it Chandra}
image, and the flux quoted in Table 1 reflects a $3 \sigma$ upper limit. 
Note that due to the northern hotspot's
off-center location, the sensitivity of {\it Chandra} was reduced by about 60\%
at this position.  

We extracted an X-ray spectrum for the southern hotspot of 3C 445, using 
\emph{specextract} in CIAO.  In order to create the spectra of 
the AGN, we defined two regions in ds9:  the hot spot and a
background region that was free from other sources.  This allowed us to create a
series of files, including source and background
PI spectra, weighted ARF, and RMF files, FEF weight files and a grouped
spectrum.  The energy range was left unrestricted for both
source spectra.  
Following this step, XSPEC was used to fit both the spectra of the hot
spot and the AGN.   
Here we discuss the spectral fits and broadband spectrum of the southern hotspot
only, as the AGN was off-center for these observations and our results for it
are fully consistent with those of Sambruna et al. (2007) and Grandi et al.
(2007).

The southern hot spot's X-ray spectrum 
was easily modeled by a basic power law with fixed
galactic absorption N$_{H}^{Gal} = 5.33\times 10^{20} ~{\rm cm^{-2}}$ ,
a photon index $\Gamma = 1.95^{+0.38}_{-0.34}$ and a $\chi^{2}$
value of 0.85, and is shown in Figure 2.  

\begin{figure}[h] \includegraphics[scale=0.35,angle=270]{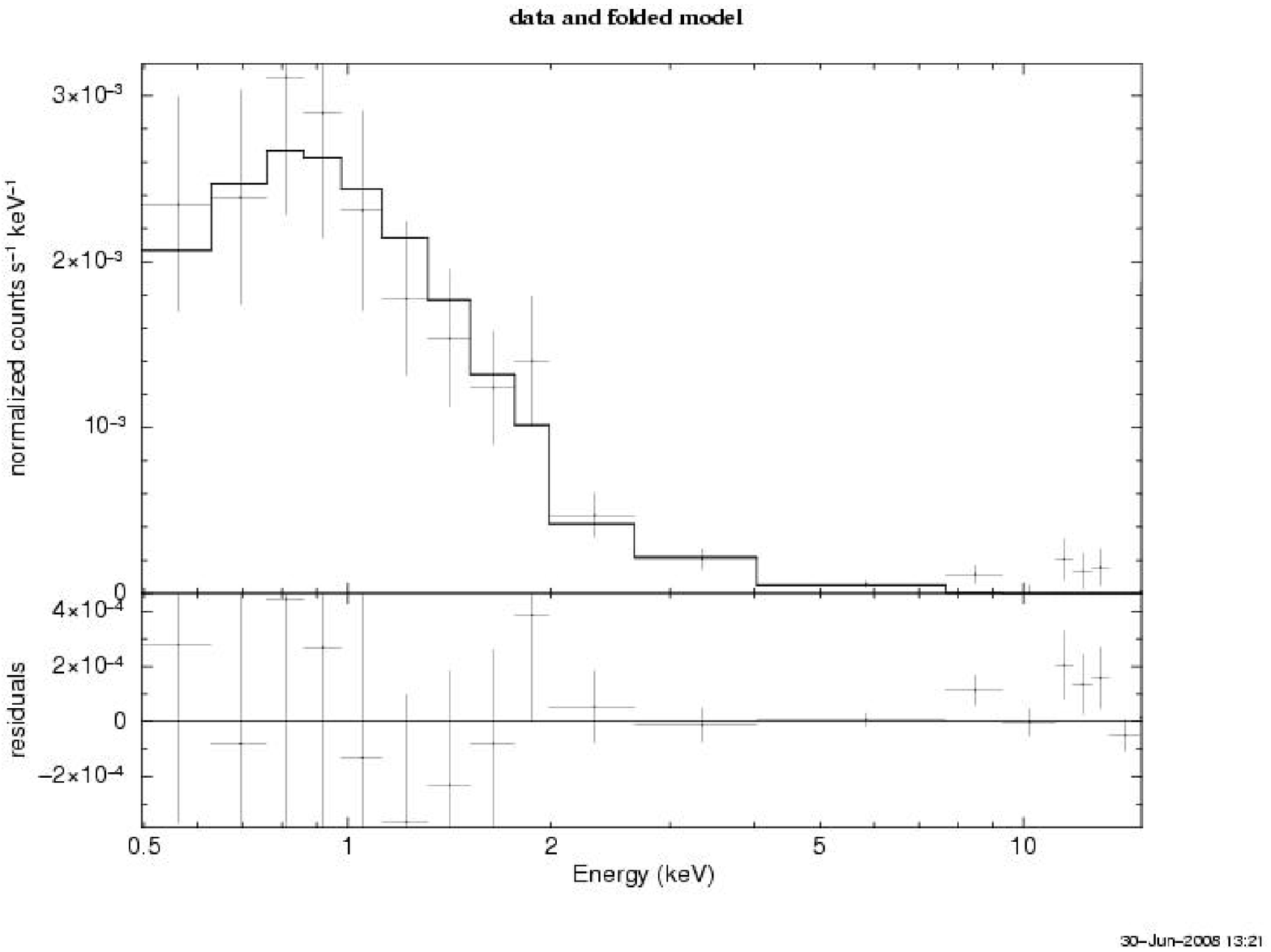}
\caption{The X-ray spectrum of the southern hot spot with the best fit power 
law overlayed.  Below it are the residuals. \label{fig2}}\end{figure}

\section{Discussion \label{interpretation}}

One of the key results of our {\sl Chandra}  observations is  that  
the X-ray emission of the southern hotspot has a very different morphology
than that seen in the radio and optical and also shows a likely displacement.
{\sl This displacement rules out all forms of one-zone models.}  Before
discussing the possible interpretations for the X-ray emission, we turn to the
radio and optical emission that are   approximately  cospatial. 
The radio-optical SED of the southern and northern hotspots are  plotted in
Figure \ref{fig:sed}, along with the X-ray points. 
The southern hotspot is brighter in the radio and optical 
by a factor of $\approx 
3$. If we attribute the brightness difference to beaming, we are forced to
conclude that the beaming of the radio-optical emitting plasma  is mild.
Below, we discuss the multiwavelength emissions of both hotspots and 
model possible X-ray emission mechanisms. 

\subsection{A  high value of $\gamma_{min}$ in the radio-optical hotspot?}  

The southern hotspot's radio--optical SED can be modeled as synchrotron emission
from a population of relativistic electrons in energy equipartition with the
hotspot magnetic field.  Assuming that beaming is not important for the 
radio--optical emission, the equipartition magnetic field is 
\begin{equation}
\displaystyle
B_{eq}=\left[{96 \pi^2  m_e c  L_{r} \nu_r^{(s-1)/2} (\gamma_{min}^{2-s}-\gamma_{max}^{2-s})\over c_1^{(s-3)/2} \sigma_\tau  V (s-2)}\right]^{2/(s+5)},
\label{eq:equip}
\end{equation} 
where $L_r$ is the radio luminosity at frequency $\nu_r$, $m_e$ is the electron
mass, $c$ is the speed of light, $\sigma_\tau$ is the Thomson cross section, and
$V$ is the volume of the emitting region. The injected electron distribution is
a power law of index $s=2\alpha_r+1=2.6$ from Lorentz factor $\gamma_{min}$ to
$\gamma_{max}$, with  $\alpha_{ro}=0.9=\alpha_r$ being the radio-optical spectral index \citep{mack09}.
To derive  equation (\ref{eq:equip}), we  assumed that an
electron  of Lorentz factor $\gamma$ in a magnetic field $B$ radiates most of
its energy at the characteristic frequency $\nu=c_1 B \gamma^2$, with 
$c_1=e/(2\pi m_e c)$. In equation (\ref{eq:equip}), $\gamma_{max}$ can in many
cases be determined observationally from the maximum observed synchrotron
frequency. However, $\gamma_{min}$ is customarily set to a value chosen by
hand. As we discuss now, there is a way to  determine observationally, or at
least constrain the value of $\gamma_{min}$, and through this get a more
appropriate value for $B_{eq}$. This in turn affects significantly the level of
both the SSC and the EC/CMB emission.

Because in this case $s>2$ and
the synchrotron emission extends for at least six decades in
frequency, $\gamma_{max}/\gamma_{min}\gg 1$, to a very good approximation
$B_{eq}\propto \gamma_{min}^{-2(s-2)/(s+5)}=\gamma_{min}^{-1.2/7.6}$: an
increase in $\gamma_{min}$ results in a mild decrease of $B_{eq}$. There are two
observational constraints on $\gamma_{min}$. An upper limit on $\gamma_{min}$ is
derived from the fact that it has to be low enough to produce the lowest
observed radio frequency from the hotspot $\nu_{r, min}  > c_1 B_{eq}
\gamma_{min}^2$.  A lower limit comes from the fact that there is no sign of
radiative cooling in the radio-optical SED, because the optical flux level is found
practically on the extrapolation of the   radio spectrum. This sets an upper limit on the
magnetic field in the radio-optical hotspot (regardless of equipartition
arguments), which in turn sets a lower limit on $\gamma_{min}$. 

\begin{figure}
\epsscale{1.10}
\plotone{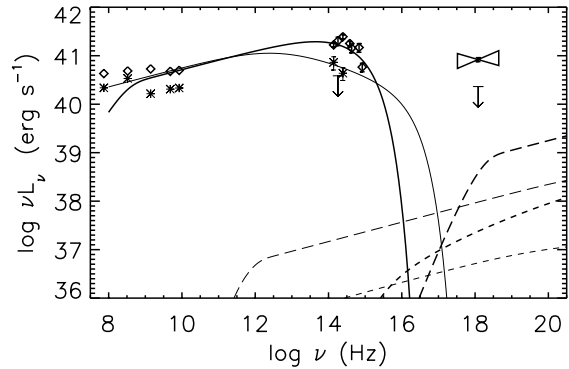}
\caption{The SED of the southern  hotspot of 3C 445 is shown with  diamonds for the radio and optical and bow-tie for the X-rays. The SED of the northern hotspot is also plotted with asterisks, including the  upper limit for the X-ray flux. The data used are taken from Table 1. Due to angular resolution constraints, the three lowest radio frequencies may include lobe emission and should be considered upper limits for the hotspot fluxes. The thin (thick) lines represent emission in equipartition conditions, assuming $\gamma_{min}=1$ ($\gamma_{min}=1840$). Solid lines represent the synchrotron, short dash lines the SSC and long dash lines the EC/CBM emission.}
\label{fig:sed}
\end{figure}

To demonstrate these considerations, in Figure \ref{fig:sed} we plot with a thin
solid line the synchrotron emission in the case of $\gamma_{min}=1$. This
corresponds to  an equipartition magnetic field, $B_{eq}=70.4\; \mu$G. 
As  can be seen, while the synchrotron SED
clearly extends below the lowest radio frequency safely associated with the
hotspot [this is the 4.8 GHz point,  because the 1.4 GHz point may be
contaminated with non-hotspot emission, (see \cite{mack09, prieto02}) as is
also true for the lower frequency points (Kassim et al. 2007).], the
synchrotron spectrum breaks at $\sim 10^{12-13}$ Hz and by doing so,
underproduces the optical emission of the hotspot. This is because the high
value of $B_{eq}$ causes a break in the electron energy distribution due to
radiative cooling (a cooling break is expected at $\gamma_b=3m_e c^2 /[4
\sigma_\tau (B^2/8\pi+U_{CMB}) R]$, where $U_{CMB}$ is the energy density of the
cosmic microwave background and $R$ is the size of the  hotspot that determines
the electron escape time $R/c$). To fit the observed SED, we need significantly 
higher values of $\gamma_{min}$. A value of $\gamma_{min}\approx 1840$ similar
to the proton to electron mass ratio $m_p/m_e$ is required to ensure that  there
is no cooling break signature at frequencies lower than optical. The value of
the corresponding equipartition magnetic field is  $B_{eq}=21.5 \; \mu$G. We
plot the resulting synchrotron SED in Figure  \ref{fig:sed} with a thick solid
line. Note that at low frequencies the model with $\gamma_{min}\approx 1840$
(thick solid line)  exhibits a break due to the high value of $\gamma_{min}$
(the slope below the break is due to the $\nu^{1/3}$ synchrotron emissivity of
electrons with Lorentz factor $\gamma_{min}$). Note also that this model manages
to  reproduce the optical emission of the hotspot.  

There is practically little freedom for $\gamma_{min}$ around $m_p/m_e$  if we
want to model the radio to optical SED with synchrotron in equipartition. Let us
mention that observationally driven arguments for similarly high  values of
$\gamma_{min}$ in hotspots of other radio galaxies have been presented by other
astronomers (e.g. Blundell et al. 2006, Stawarz et al. 2007, Godfrey et al.
2009).  However, some jet sources require much lower values, e.g., PKS 0637--752
(Mehta et al. 2009). Values of $\gamma_{min}\approx m_p/m_e$ are particularly
interesting, because this is the minimum energy that electrons crossing a shock
must have to be picked up efficiently by Fermi acceleration (e.g. Spitkovsky
2008). The level of both the EC and SSC emission depend on the value of
$B_{eq}$, which in turns depends on $\gamma_{min}$:  because $L_{SSC, EC}/L_S
\propto U_B^{-1}\propto B^{-2}$, and $B_{eq}\propto
\gamma_{min}^{-2(s-2)/(s+5)}$, for $B=B_{eq}$, $L_{SSC, EC} \propto 
\gamma_{min}^{4(s-2)/(s+5)}=\gamma_{min}^{2.4/7.6}$. Therefore, an  increase
from $\gamma_{min}=1$ to   $\gamma_{min}=1840$, should increase the EC and SSC
level by a factor of $\approx 10.7$, as seen in Figure \ref{fig:sed}. Even for
$\gamma_{min}=1840$,  the X-ray emission of the radio-optical hotspot is much
weaker than the upstream detected  X-ray emission.


If analyses like the above point toward a high value of $\gamma_{min}$, then
future low frequency-high angular 
resolution observations ({\it e.g.}, with the
LWA) should detect the SED break at low frequencies. Existing low fequency 
radio observations suggest such a break for the eastern hotspot of Cygnus A
\citep{lazio06}. For 3C 445,  VLA observations at 74 MHz and 330 MHz (Kassim et
al. 2007),  provide us with upper limits for the  emission of the hotspots (due
to lobe contamination). The low frequency data plotted in figure \ref{fig:sed}
may be contaminated by lobe emissions and hence
are not sufficient to constrain the spectral shape below 4.8 GHz.  
If a  low frequency
break is found by future low frequency-high resolution observations, they
will strengthen the above picture. If on the other hand the low frequency 
radio spectrum exhibits
no such break, we will have to search for an alternative reason for the lack of
cooling break (as we discussed above, a low value for
$\gamma_{min}$, manifested through a lack of low frequency break, requires 
a higher value for $B_{eq}$, which in turns
produces a cooling break below the optical, as can be seen from the thin solid
line in  Figure \ref{fig:sed}). Distributed reacceleration is a very plausible
candidate and it has been claimed to explain the optical emission of 3C 445
(e.g. Prieto et al. 2002). 

\subsection{What is the X-ray emission mechanism? \label{x}}

Any interpretation of the X-rays must take into account ($i$) the level and
spectrum of the X-ray emission,  ($ii$) the apparent 
upstream `nesting'  of the X-ray emission into the hollow part of the east-west arc of radio--optical emission, and  ($iii$)
the mostly westward displacement of the X-ray
emission relative to the peak of the radio - NIR 1  hotspot.
Also, because of  the moderate
hotspot to counterhotspot flux ratio in the radio and optical, it should not 
invoke significant beaming for the radio-optical emitting plasma. These
considerations automatically exclude the possibilities of one zone EC/CMB
\citep{tavecchio05} and SSC in equipartition, because in these models both the SSC and EC/CMB emission is
cospatial with the radio--optical hotspot emission and (see Fig. \ref{fig:sed})
is much lower than the X-ray detected flux).

\begin{figure}
\epsscale{1.1}
\plotone{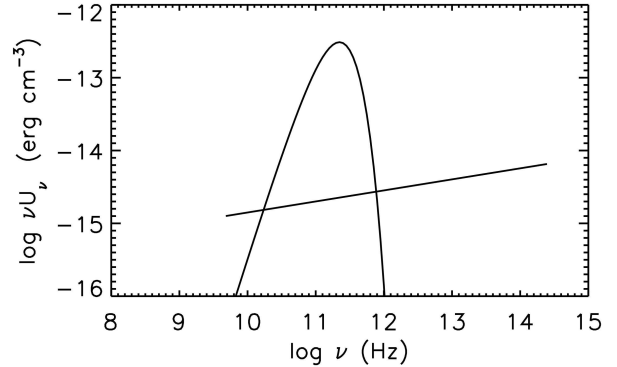}

\caption{The energy density at the location of the southern hotspot. The
straight line represents the photon energy density due to the radio
optical-emission of the hotspot, while the blackbody is the CMB.}

\label{fig:seeds}
\end{figure}

A displacement between the X-ray and radio emission is predicted in the case of 
UC emission from a decelerating flow, in which freshly accelerated relativistic
electrons from the fast base of the flow upscatter to high energies the radio
photons produced in  the downstream slow part of the flow by electrons that have
cooled radiatively \citep{georganopoulos03}. This version of the UC emission
from a decelerating flow is  not favored, however, as it predicts that 
the optical and
the X-rays are co-spatial and that the radio emission is shifted downstream.
This is because optical emission can only be produced by the fast, upstream part
of the flow, where the freshly accelerated electrons are found, the same place
from which the UC X-rays are produced. Also, the model predicts the optical
to be more beamed than the radio, something not supported by the very similar
hotspot to counter-hotspot flux ratio in radio and optical. Finally, the model
uses as seed photons the synchrotron photons produced in the source, making the
implicit assumption that these photons dominate  the local photon energy
density. This is not the case, as can be seen in Figure \ref{fig:seeds}.

The fact that the X-ray emission is displaced from the radio-optical
hotspot requires the X-ray emitting electrons to also be displaced from
the radio-optical emitting electrons. If the X-ray emission is of IC nature,
these electrons will experience the seed photon field of Figure \ref{fig:seeds},
provided they are located at a distance from the radio-optical hostspot not much
greater than the radio-optical hotspot size.  This immediately  excludes the
IR-optical photons as seed photons for the X-ray emission, because this would
require the presence of a powerful component of IC emission due to CMB photons 
upscattered in $\sim$ the optical band,  co-located with the X-ray component.
This is not observed. We therefore reach the conclusion that {\sl if the X-ray
emission is due to a particle  population located upstream of those in other
bands, then the seed photons that these electrons upscatter must be the CMB. }

A model that produces co-spatial radio-optical synchrotron emission and EC/CMB
X-ray emission shifted upstream, has been proposed by \cite{georganopoulos04} to
address such  displacements observed in the large scale jets of quasars. The
model assumes that distributed particle acceleration offsets radiative losses.
In this model,  a relativistic decelerating flow results in an increase of the
magnetic field and electron density at the slow downstream part of the flow,
increasing  the synchrotron emissivity. At the same time, the faster upstream
part of the flow experiences a higher CMB comoving energy density
$U_{CMB}\propto \Gamma^2$, where  $\Gamma$ is the bulk Lorentz factor of the
flow (see Figure 3 of Georganopoulos \& Kazanas 2004).  In this scenario
the only displacement observed is along the flow axis.
 In our case, however, besides the general upstream shift of the X-rays relative to the radio--NIR emission, we note that the peak of the radio--NIR 1 emission is shifted  by $\sim 2" $ to
the east relative to the center of the X-rays ($1"$ corresponds to $1.07$ Kpc). Therefore, for
this model to still be viable, the flow needs to bend to the east after
producing the X-ray emission.

\begin{figure}
\epsscale{1.10}
\plotone{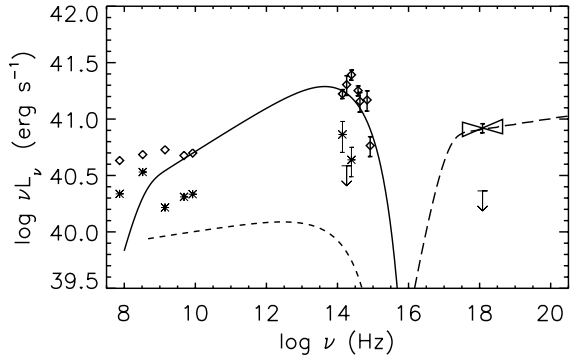}
\caption{  X-rays due to EC/CMB (long dash line). The emission is assumed to come
from plasma moving relativistically (bulk Lorentz factor $\Gamma=4$) at an angle $\theta=14^\circ$ to the line of sight. To reproduce the observed displacements we assume that the flow decelerates and  bends to $\sim \theta=30^\circ$ and terminates $\sim 4.6$ Kpc downstream, producing the radio optical emission (solid line). The radio-optical emission coming from the X-ray spot is much weaker (short dash line). The data points are the same plotted in Figure \ref{fig:sed}.}
\label{fig:sedX}
\end{figure}

To demonstrate how this  could reproduce the observed upstream X-ray emission,
we plot in Figure \ref{fig:sedX}  the emission that would result from plasma in
equipartition moving with a bulk Lorentz factor $\Gamma=4$ at an angle to the
line of sight $\theta=14^\circ$ and carrying the same power as that injected in
the radio-optical hotspot ($L_{kin}=1.3 \times 10^{44} $ erg s$^{-1}$). 
Although such a small angle to the line of sight, bordering angles 
typical of blazars is admittedly uncomfortable, it cannot be excluded for 
this broad line FR II radio galaxy.
To reproduce the shift of the radio-optical emission in NIR 1 relative to
location of the X-ray emission, the flow must bend to the east,
forming an angle of $30^\circ$ to the line of sight. The projected physical
distance between the X-ray and radio-optical components is 4.6 kpc. Bends in the
flow and velocity gradients at the hotspots, such as the one suggested here,
are  seen in numerical simulations of relativistic flows (e.g.  Aloy et al.
1999).  Deceleration from $\Gamma=4$ to the subrelativistic speed of the
radio-optical hotspot could be achieved by a series of oblique shocks that could
also aid electron re-acceleration. This latter possibility could be tested by
future, high-resolution radio imaging of the hotspot. If indeed this is the
X-ray emission mechanism, we expect that due to relativistic dimming, the X-ray
emission from the counter hotspot would be undetectable, even for very deep 
{\sl Chandra} exposures. On the assumption of hotspot-counter hotspot symmetry,
X-ray detection of the counter hotspot would exclude this last alternative for
an inverse Compton interpretation of the X-rays.

An alternative model, which avoids
the uncomfortably small jet angle to the line of sight and the relatively high Lorentz factor
of the X-ray emitting plasma needed for the above interpretation, together with
the arguments against strong beaming in other sources (see \S \ref{intro}), is
to generate the X-ray emission via synchrotron radiation from a
high energy population ($\gamma\sim 10^8$) of electrons accelerated upstream of
the radio-optical hotspot.  A possible way to produce this electron population
upstream the radio-optical emitting region is through acceleration in the
reverse shock of a reverse-forward shock structure (in this picture the
radio-optical comes from electron acceleration in the forward shock; Kataoka et
al. (2008) proposed it, motivated by the upstream, relative to the radio, 
X-ray emission seen in both hotspots and jets of 3C  353).  This scheme, 
however, as Kataoka et al. (2008) note,  does not address the reason the two 
shocks accelerate electrons at different energies, with the reverse shock 
consistently reaching higher electron energies.   If the X-ray emission of the southern hotspot is synchrotron, then
the observed emission in the {\it Chandra} band must lie close to 
the peak of its $\nu f_\nu$ emission. This is because the X-ray photon 
index is $\sim 2$ and the luminosity of this component at optical
energies must  be below the level of the optical emission detected downstream.
The emission could in principle extend to energies higher than the {\it
Chandra} band but this would require the presence of unrealistically energetic
electrons.) 
Because cooling of the X-ray emitting electrons is severe, even if we only
consider the CMB photon energy density, these electrons are suffering strong
cooling, which must be  balanced by continuous reacceleration.   

Finally, an account of most of the phenomenology may rest with the possibility 
that the entire radio-NIR-X-ray emission is synchrotron, and that the
jet beam moves laterally with time to the west, implying that
the maximum X-ray emission is the most recent and naturally further displaced
from the AGN core. For this same reason (age) there is 
less IR and almost no radio
emission in along this direction (the corresponding electron radiative times 
are longer). The fact that the radio emitting electrons have the longest 
radiative lifetime would then explain the displacement of the maximum emission 
at this frequency (and of the IR) to the East  and the absence of X-rays in the 
same region (the X-ray emitting electrons have all cooled to lower energies).
There is some indication of such a motion from the fact that the AGN core and 
the two lobes do not all line on a straight line. 

To conclude, both the EC/CMB from a decelerating flow  and synchrotron 
interpretation of the X-rays face important problems, although the EC/CMB 
model is more constrained and, therefore, easier to falsify, while the 
synchrotron interpretation  can reproduce any observed emission by introducing 
additional electron populations as needed.
A purely spectral discrimination between the synchrotron and inverse-Compton 
models is not possible, as both can be made compatible with various X-ray 
slopes as well as the forms for the 'valley' in between the X-ray and radio-IR
components (see
e.g., Uchiyama et al. 2006, 2007; Jester et al. 2007; and  Hardcastle et al.
2004).  The synchrotron interpretation, however, does not require beaming and 
is the only alternative among those examined that could produce detectable X-ray
emission  from the counter-hotspot. Therefore, detection of bright X-ray
emission from the counter hotspot in
future X-ray observations would rule out any reasonable IC
models for the hotspots of 3C 445. An additional test of the nature of the X-ray
emission could come from future X-ray polarimeters like GEMS: while  the
synchrotron emission is expected to be highly polarized, the EC/CMB should
produce negligible polarization (Begelman \& Sikora 1987,
McNamara et al. 2009, Uchiyama \& Coppi in prep.). An approach that can 
yield results with our current observational capabilities is based on HST UV 
polarimetry:
if the far UV emission is shown to be the low energy tail of the X-ray 
component as is the case with 3C 273 (e.g. Jester et al. 2007), then UV HST 
imaging polarimetry of the hotspots  will distinguish between the EC/CMB and 
synchrotron mechanisms (e.g., McNamara et al. 2009, Uchiyama \& Coppi, in
prep.).

\acknowledgments

We thank an anonymous referee for comments that significantly strengthened
this paper. 
This work was supported at FIT and UMBC by the Chandra grant  G07-8113A and the
NASA LTSA grant NNX07AM17G.  This project was also partially funded by a
partnership between the  National Science Foundation (NSF AST-0552798), Research
Experiences for Undergraduates (REU), and the Department of Defense (DoD) ASSURE
(Awards to Stimulate and Support Undergraduate Research Experiences) programs.

\pagebreak

\end{document}